\documentclass[1--150]{article}
\usepackage{url}
\usepackage[pdftex]{graphicx}
\usepackage{amssymb}

\begin{document}

\title{Status and New Ideas Regarding Liquid Argon Detectors \footnote{Accepted for publication in Annual Review of Nuclear and Particle Science}}

\author{Alberto Marchionni \\
Fermi National Accelerator Laboratory, Batavia, Illinois 60510 \footnote {Operated by Fermi Research Alliance, LLC under Contract No. DE-AC02-07CH11359 with the United States Department of Energy.} \\email: alberto@fnal.gov}

\maketitle


\begin{abstract}
Large (up to $\sim 100$ kt) liquid argon time-projection chamber detectors
are presently being considered for proton decay searches and neutrino astrophysics,
as well as for far detectors for the next generation of long-baseline
neutrino oscillation experiments that aim to determine neutrino mass hierarchy
and search for CP violation in the leptonic sector. These detectors
rely on the capabilities to assemble large volumes of LAr in ultrahigh-purity
conditions, possibly in an underground environment, and to achieve relatively
long drifts for the ionization charge. Several proposals have been
developed, each of which takes a different approach to the design of the
cryogenic vessels and has different scales of modularity to reach the final
mass dictated by physics. New detector concepts, with innovative designs
of readout electronics and novel methods for the readout of the ionization
charge and scintillation light, have been proposed.
\end{abstract}


\section{INTRODUCTION}
\label{sec:intro}
The use of a liquid argon (LAr) ionization chamber~\cite{marshall:232} was first demonstrated in the early 1950s, but it was only at the end of 1960s, following a suggestion by Alvarez~\cite{Alvarez:1968}, that liquid noble gas devices were considered as high energy particle detectors.

In 1977, Rubbia~\cite{Rubbia:1977} proposed a liquid argon Time Projection Chamber (LAr TPC) for neutrino detection. Other groups~\cite{Chen:1978yh,Gatti:1979} were developing the idea of a LAr TPC at almost the same time. 

Thanks to the high mobility~\cite{Shibamura1975249} and low diffusion~\cite{Derenzo:1974ji} of electrons in liquid Ar, large LAr volumes can be operated as TPCs, providing high quality imaging and high resolution energy measurements from the detection of the ionization charge~\cite{Miyajima:1974zz,Miyajima:1974zy}. The relatively high density of LAr and its commercial availability enable one to assemble large masses of argon suitable for use as both the target and the detector of incoming particles. A LAr TPC is a continuously sensitive detector that can measure, in addition to the ionization charge, the scintillation light~\cite{jortner:4250,Hitachi:1983zz,Doke1990617} emitted by excited argon diatomic molecules (excimers) to determine the absolute event time and to provide a self-trigger with no bias on the
charge detection.

Large LAr volumes can be readout~\cite{Gatti:1979} with high spatial granularity, of the order of a few cubic millimeters, by finely instrumenting only two of the three spatial coordinates and, following the TPC principle, using the drift time to measure the third coordinate.

The performance of a LAr TPC critically depends on the amount of electronegative impurities~\cite{Swan196374,Hofmann:1976de,Bakale1976}; impurity concentrations less than a few tens of parts 
per trillion are required for drift paths of several meters. High argon-purification rates, which are necessary in all large LAr detectors, can be achieved with argon purification in liquid phase, 
as first discussed in Reference~\cite{Cennini1993567}.

LAr TPCs overcome the deficiencies of both bubble chambers (which are limited in size and sensitive only for short times) and of large size calorimetric detectors (which suffer from coarse granularity and limitations in the identification of electromagnetic showers). LAr TPCs, thanks to their high readout granularity, can sample electromagnetic showers down to a few percent of a radiation length, making them excellent detectors for electron identification. When compared with Water Cerenkov detectors, LAr TPCs permit lower momentum thresholds for the identification of particles heavier than electrons, notably protons, and are predicted to provide greater electron identification efficiency in presence of a $\pi^{0}$ background.

High-energy physicists quickly realized the potential to use LAr TPCs to detect neutrinos from astrophysical sources, in particular solar neutrinos~\cite{Bahcall1986324}, 
and more generally to detect very rare events such as proton decay~\cite{Giboni:1984uq}. Since the mid 1980s, the ICARUS Collaboration, led by Rubbia, has submitted 
a series of proposals~\cite{ICARUS85,ICARUS89,ICARUS93} to the Gran Sasso Laboratory (LNGS) in Italy for the underground installation of kilotonne-scale LAr TPC modules; 
at the same time, they initiated a dedicated research and development (R\&D) program. 

The ICARUS Collaboration is currently operating the largest LAr TPC ever built, the ICARUS T600~\cite{Amerio:2004ze,Rubbia:2011ft}. This instrument, located underground at LNGS 
and operated in conjunction with the CERN to Gran Sasso neutrino beam, has an active mass of $\sim500$ t, demonstrating the feasibility of using a large LAr installation
in an underground environment.

Detectors ~\cite{Cline:2001pt,Rubbia:2004tz,Bartoszek:2004si,Cline:2006st,Baibussinov:2007ea,Rubbia:2009md,Akiri:2011dv,Stahl2012xx}
from 10 kt scale up to 100 kt have been proposed for future experimental programs that would combine proton decay searches, observation of atmospheric and supernova neutrinos, 
and long baseline neutrino oscillation physics. Such programs would optimally exploit the features of LAr TPC detectors, which are ideally suited for making measurements 
in the MeV to GeV range and demonstrate superior performance in electron identification. Clean electron measurements are crucial in long-baseline neutrino oscillation experiments, with
conventional wide-band neutrino beams, for measurements of the $\nu_\mu \longrightarrow \nu_e$ transition probability, the determination of the neutrino mass ordering, and especially 
searches for CP violation in the leptonic sector~\cite{Diwan:2003bp}.

A 10 kt scale LAr detector, if installed underground even at moderate depth, would extend the search for proton decay via modes favored by supersymmetric grand unified models 
(e.g. $p \rightarrow K^{+} \bar{\nu}$) above $\sim$10$^{34}$ years. Such a detector would have between 5 and 10 times the efficiency
of water Cherenkov detectors for such decays~\cite{Bueno:2007um}. 

Core collapse supernova neutrinos can be detected by LAr TPCs through different channels~\cite{GilBotella:2003sz}, namely elastic scattering on atomic electrons and neutral and charged current processes on
argon according to the neutrino flavor. The possibility to observe supernova relic neutrinos was considered in Reference~\cite{Cocco:2004ac}.
Direct observation of the $\tau$ neutrino component in atmospheric neutrinos with a LAr TPC was first suggested in Reference~\cite{Rubbia:1999py} and more recently discussed in Reference~\cite{Conrad:2010mh}.
The low energy frontier in LAr, down to energies of tens of keV, is being actively explored by direct Dark Matter search experiments~\cite{Chepel:2012sj}.

LAr TPCs, possibly magnetized~\cite{Cline:2001pt,Ereditato:2005yx}, have also been proposed for use in neutrino factory beams~\cite{Bueno:2000fg,Cline:2001pt} to identify, in addition to muons, final states with electrons and $\tau$ leptons. More recently LAr TPCs have been proposed~\cite{Chen:2007zz,Antonello:2012hf} for use as detectors in short-baseline experiments for neutrino cross section measurements 
and in searches for sterile neutrinos.

A size increase of a factor 10 to 100 with respect to the present ICARUS T600, though challenging, is not unrealistic, as we argue below. 
In Section~\ref{sec:historical}, we provide a historical overview of LAr technology; Section~\ref{sec:properties} summarizes the thermal and physical properties of LAr. 
In Section~\ref{sec:RandD}, we discuss in detail the main issues in scaling from an ICARUS-like detector to a tens-of-kilotonnes device and describe ongoing R\&D. 
Different design concepts, as well as intermediate-scale detectors and test-beam setups, are reviewed in Section~\ref{sec:designs}. In Section~\ref{sec:conclusions}, we present our conclusions.

\section{HISTORICAL OVERVIEW OF LAr TECHNOLOGY}
\label{sec:historical}
Long ago, researchers recognized that ionizing radiation in LAr can excite electrons into conduction levels so that they can be transported through the liquid by an applied electric field. 
At the end of the 1940s, initial observations~\cite{Davidson48,Gerritsen1948407,Hutchinson48,Davidson50} of conductivity pulses in LAr demonstrated high
electron mobility; these observations led to calorimetric measurements of ionizing radiation~\cite{marshall:232}.

In 1968 Alvarez ~\cite{Alvarez:1968} suggested that liquid noble gases detectors could be used as high spatial resolution detectors, given the expected low diffusion of electrons in liquid, 
and as total absorption counters with good energy resolution. This proposal led to the development of a multistrip LAr ionization chamber with a better position resolution than that 
of gas detectors~\cite{Derenzo:1974ji}. This instrument also provided the first measurement of electron diffusion in LAr.

At the end of the 1960's, Doke (see Reference~\cite{Doke:1993nc} for a historical review of early R\&D on noble liquids) began taking comprehensive measurements 
of the fundamental properties of liquid Ar and liquid Xe as detector media. At approximately the same time, to obviate to the unsuccessful attempts to achieve charge amplification 
in liquid argon~\cite{Derenzo:1900zza} and still obtain relatively large signals, Dolgoshein et al.~\cite{Dolgoshein:1970,Dolgoshein:1973} proposed and successfully tested 
the main features of double-phase argon detectors, in which ionization electrons produced in the liquid are extracted to the argon vapor on top and are there amplified. 

In 1974, aiming to create a high energy particle detector with good energy resolution, Willis \& Radeka~\cite{Willis:1974gi} built and operated a LAr ionization chamber 
with a large number of thin metallic plates acting as readout electrodes. They employed specifically developed low noise electronics that were optimized for large capacitances. 
Rubbia's~\cite{Rubbia:1977} 1977 proposal of a LAr TPC gave rise to the concept of a detector that could act simultaneously as a homogeneous calorimeter and as a
tracking device and would have an imaging quality comparable to that of a bubble chamber.

In a TPC, ionization electrons drift from a cathode towards a readout structure by means of a uniform electric field~\cite{Blum:2008zza} to achieve a constant drift velocity. 
The electric field is generated by surrounding the drift volume by a set of equally spaced electrodes, denoted field shapers, that are kept at distinct potentials 
such that the difference in voltage between two contiguous electrodes is constant.

Reference~\cite{Gatti:1979} describes the standard method for reading out the ionization charge in LAr TPCs. 
This method, pioneered by the ICARUS Collaboration, employs three parallel wire planes immersed in LAr, with wires at different angles. 
By properly biasing the wire planes, the first two wire planes become fully transparent to the drifting electrons and can detect a signal
induced by the moving electrons, whereas the last plane acts as a collecting electrode. 
The signals (see Reference~\cite{Amerio:2004ze} for a sketch of the signal shapes) are digitized at a frequency of a few megahertz.
Each wire plane provides a two-dimensional view of the event; one coordinate is given by the wire number and the other by the drift time.

\subsection{LAr purification techniques}
The LAr TPC concept requires drift paths of at least several tens of centimeters. This constraint imposes strict limits on the concentration of electronegative impurities in LAr (see Section~\ref{sec:purity}).

Two experimental groups initiated investigations into argon purification systems and long drift lengths. The Chen group~\cite{Chen:1978yh,Chen1981454} first achieved 
electron attenuation lengths of $\sim35$ cm with drift fields of 1 kV/cm. A large scale argon purification system~\cite{Doe1987170} employed gas and liquid phase purification 
by using reduced copper and a mixture of molecular sieves (4A, 5A and 13 X), respectively, in a closed loop recirculation of the evaporated argon. 
This system attained  attenuation lengths $\ge3.7$ m at drift fields of 0.5 kV/cm, which correspond to an impurity level of $\le0.2$ ppb oxygen equivalent.

The Rubbia group~\cite{Aprile:1985xz,Buckley:1988qx} investigated electron lifetimes in small, gridded ionization chambers, operated at very low drift fields, by analyzing signal pulse shapes. 
The purification system, operated in the gas phase, employed a mixture of molecular sieves and silica gel to remove water, carbon dioxide and hydrocarbons, 
along with a commercial Oxisorb~\cite{oxisorb} getter, made of reduced chromium oxide on a silica gel carrier, to absorb oxygen. 
A 10 cm drift chamber was operated~\cite{Aprile:1985xz} at drift fields down to 10 V/cm, with drift times of 1.8 ms. 

Subsequently, a chamber with 153 mm drift gap and drift fields as low as 5 V/cm attained electron lifetimes of $\sim9$ ms~\cite{Buckley:1988qx}, which correspond 
to an impurity concentration of 0.03 ppb oxygen equivalent. This development led to the creation of electron lifetime detectors~\cite{Carugno:1990kd}. These
detectors are based on attenuation measurements of electrons, extracted by a Nd:YAG UV laser beam from a metallic photocathode, in a double-gridded drift chamber. 

The ability to purify LAr directly in liquid phase~\cite{Cennini1993567} by use of a combination of molecular sieves and Oxisorb cartridges
demonstrated the feasibility of employing the ultrapure LAr technique over very large volumes. 

\subsection{Development of LAr TPC detectors}
\label{sec:TPCdetector}
The bidimensional tracking capabilities of a LAr TPC were first demonstrated by Chen and colleagues~\cite{Doe1982639}, who used 
drift distances of up to 25 cm and a multistrip readout anode with a 2.5 mm pitch.

Increasingly large detectors have been constructed and tested by the ICARUS Collaboration over the past 20 years. 
A two-dimensional LAr TPC~\cite{Bonetti:1989yi} with a 24 cm drift gap, the first to be equipped with a sense-wire (induction) 
plane to demonstrate nondestructive detection of the electron image~\cite{Gatti:1979} produced by ionizing events, was exposed 
to a 5 GeV pion beam at the CERN PS and used to study low energy electrons by observations of the associated $\delta$ rays.

For the first time, a 3 t LAr TPC~\cite{Benetti:1993yn,Cennini:1994ha}, with a maximum drift distance of 42 cm, produced
three-dimensional images of ionizing events over a volume of $\sim1.5$ m$^3$. A screening grid was followed by two 
2 mm pitch wire planes that worked in induction mode and collection mode. Given the small ionization charge signal released in 
LAr by a minimum ionizing particle (MIP) (see Section~\ref{sec:ioniz}), the use of low-noise readout electronics is required 
(see Section~\ref{sec:wireread}). This 3 t LAr TPC employed charge-sensitive preamplifiers with a junction field effect transistor (JFET)
input stage in an unfolded cascade architecture~\cite{Radeka197451}.

In the 1990s, the ICARUS-Milano Collaboration pioneered the use of LAr TPCs as neutrino detectors with the exposure 
of a 50 liter chamber~\cite{Arneodo:2006ug} on the CERN West Area Neutrino Facility. The front-end electronics~\cite{Baibussinov:2011zw} 
were based on JFETs operated in LAr close to the readout wires, which minimized the preamplifier input capacitance due 
to the absence of long readout cables.

The detection of scintillation light~\cite{Cennini:1999ih} in coincidence with ionizing tracks in a LAr TPC can be used as 
an absolute time measurement  for the spatial localization of the ionizing event. Conventional glass-window photomultiplier 
tubes (PMTs) are not sensitive to the  far vacuum ultraviolet (VUV) light produced in LAr, which would require the use of costly PMTs 
with MgF$_2$ windows. A suitable solution~\cite{Benetti200389} makes use of glass window PMTs immersed in LAr; 
these PMTs were developed specifically for use at cryogenic temperatures and are coated with a fluorescent wavelength 
shifter~\cite{Burton:73} to the PMT sensitive region.

The construction of the ICARUS T600 detector~\cite{Amerio:2004ze} was an important step in detector size. This detector is composed of a large cryostat split into two identical, adjacent 300 t modules, each of which has an internal volume of $3.6 \times 3.9 \times 19.6$ m$^3$. The ionization charge is detected by wire chambers and scintillation light by an array of 74 PMTs~\cite{Ankowski:2006xx}. The TPC wire chambers are positioned on the opposite sides of a cathode plane placed in the middle of each 300 t module. Each wire chamber consists of three parallel wire planes, with 3 mm wire pitch and wires at $0^\circ$ and $\pm60^\circ$, with respect to the horizontal direction; the wires are readout by low noise warm electronics located outside the cryostat (see Section~\ref{sec:wireread}). With a maximum drift distance of $\sim1.5$ m, a cathode voltage of 75 kV is required for a nominal electric field of 500 V/cm. A cryogenic, custom-made, high voltage (HV) feedthrough carries the HV generated by an external power supply to the inside of the cryostat on the cathode plane. It successfully operated up to a maximum of 150 kV~\cite{Amerio:2004ze}. 

Argon purification is accomplished by both gas and LAr recirculation through Oxisorb~\cite{oxisorb} getters and molecular sieves. For each 300 t module, the liquid is recirculated by means of an immersed cryogenic pump placed inside an independent dewar. This setup recirculates the full volume in 6 days.

A 300 t module was first operated on the surface; it recorded cosmic ray data for a period of about two months. The electron lifetime reached a value of 1.8 ms and was still increasing by the end of the run~\cite{Amoruso:2004ti}. 

The succesful installation of the ICARUS T600 detector~\cite{Rubbia:2011ft} in the LNGS underground laboratory was a major milestone in LAr TPC technology. The detector is presently collecting data with a good electron lifetime, of up to 7 ms, and it is being used to study neutrino physics~\cite{Antonello:2011aa,Antonello:2012hg,Antonello:2012pq}.

\section{LAr PROPERTIES}
\label{sec:properties}
Argon, produced industrially by the fractional distillation of liquid air, is readily commercially available, since it constitutes 1.28\% by weight of air. Annual US production alone amounts to about 1000 kt/year. 

The main isotopes of argon~\cite{NISTElem} are $^{40}$Ar (99.6\%), $^{36}$Ar (0.34\%), and $^{38}$Ar (0.06\%), all of which are stable against radioactive decays. The long-lived radioactive isotope $^{39}$Ar, a beta-emitter with $T_{1/2}=269$ years and a Q-value of 565 keV, is also present in atmospheric argon, produced by cosmic rays via the $^{40}$Ar(n, 2n)$^{39}$Ar process. 
Measurements in a two-phase argon chamber~\cite{Benetti:2006az} result in a $^{39}$Ar specific activity of $(1.01 \pm 0.08)$ Bq/kg of natural argon, consistent with a previous determination~\cite{Loosli198351} in atmospheric argon. 

 Tables~\ref{TabLAr_th} and~\ref{TabLAr_ph} present the main thermal and physical properties of LAr.
In the following subsections, we briefly discuss the relevant physical processes for a LAr TPC.

\begin{table}
\begin{minipage} {\textwidth}
\caption[Caption for LOF]%
{\label{TabLAr_th} Thermal properties of LAr~\cite{NISTWebBook,tegeler:779}.\footnote{NBP: normal boiling point}}
\begin{tabular} {@{}lccc@{}}
\hline
Atomic number & 18\\
Atomic mass & 39.948 u\\
\hline
 & Pressure (kPa) & Temperature (K) & Density (g/cm$^3$) \\
\hline\hline
Triple point & $68.891 \pm 0.002$ & 83.8058 &  \\
Liquid phase & 100.0 & 85 & 1.410 \\
Saturation properties& 88.110 & 86 & 1.403 (liquid)\\
                &           &     & $5.080 \times 10^{-3}$ (vapor) \\
\multicolumn{1}{r}{(NBP)} & 101.325 & $87.303 \pm 0.002$ & 1.395 (liquid)\\
                &           &                  &  $5.774 \times 10^{-3}$ (vapor)   \\
                & 109.01 & 88 & 1.391 (liquid)\\
                &            &      &  $6.174 \times 10^{-3}$ (vapor)) \\
Critical point & $4863 \pm 3$ & $150.687 \pm 0.015$ & $0.5356 \pm 0.0010$  \\
\hline
At NBP \\
\hline\hline
\end{tabular}
\begin{tabular} {@{}lc@{}}
Isobaric heat capacity (C$_p$) & 1.117 kJ/kg K \\
Thermal conductivity &  0.1256 W m$^{-1}$ K$^{-1}$\\
Latent heat of vaporization & 161.1 kJ/kg \\
Viscosity & 270.7 $\mu$Pa s \\
Expansion ratio liquid (NBP) to gas (1 atm, 20 $^\circ$C) & 840 vol/vol \\
\hline
\end{tabular}
\end{minipage}
\end{table}

\begin{table}
\begin{minipage} {\textwidth}
\caption[Caption for LOF]%
{\label{TabLAr_ph} Physical properties of LAr.\footnote{MIP: minimum ionizing partcle; NBP: normal boiling point}}
\begin{tabular}{@{}lc@{}}
\hline
Passage of particles & \\
\hline\hline
Stopping power $\langle dE/dx \rangle$ (MIP)~\cite{Beringer:1900zz}   & 1.508 MeV cm$^2$/g\\
Nuclear interaction length~\cite{Beringer:1900zz} & 119.7 g/cm$^2$ \\
Mass attenuation coefficient for 1 MeV $\gamma$~\cite{NISTElem} & $5.762 \times 10^{-2}$ cm$^2$/g\\
Critical energy (electrons)~\cite{Beringer:1900zz} & 32.84 MeV (for e$^-$)\\
                                                                        & 31.91 MeV (for e$^+$)\\
Radiation length~\cite{Beringer:1900zz} & 19.55 g/cm$^2$ \\
Moliere radius~\cite{Beringer:1900zz} & 12.62 g/cm$^2$ \\
W$_{i}$ (average energy per ion pair, 1 MeV e$^-$)~\cite{Miyajima:1974zz,Miyajima:1974zy} & $23.6 \pm 0.3$ eV \\
W$_{ph}$ (average energy per photon, 1 MeV e$^-$)~\cite{Doke1990617} & 25 eV\\
Peak VUV scintillation wavelength~\cite{jortner:4250,Heindl:2010zz} & $\sim127$ nm\\
Scintillation decay time constants, with & $\tau_{fast} = 7 \pm 1$ ns (23\%) \\
relative proportions for 1 MeV e$^-$~\cite{Hitachi:1983zz} & $\tau_{slow} = 1.6 \pm 0.1$ $\mu$s (77\%) \\
\hline
Electrical and optical properties & \\
\hline\hline
Ionization energy (gas)~\cite{NISTElem} & 15.7596 eV \\
Ionization energy (liquid, computed)~\cite{Badhrees:2010zz,Schmidt1984xx} & $13.6-13.8$ eV \\
Dielectric strength (gold-gold electrodes)~\cite{Swan1961448} & $\sim 750$ kV/cm \\
Static dielectric constant (NBP)~\cite{Amey:146} & 1.504\\
Refractive index (NBP, 435.8 nm)~\cite{Sinnock1969181}& 1.2325\\
Refractive index (128 nm, computed)~\cite{Seidel:2001vf} & 1.38 \\
\hline
Free electron properties & \\
\hline\hline
e$^-$ drift velocity (500 V/cm, 89 K)~\cite{Walkowiak:2000wf,Amoruso:2004ti} & 1.55 mm/$\mu$s \\
D$_L$ (drift field 100--350 V/cm)~\cite{Cennini:1994ha} &  $4.8 \pm 0.2$ cm$^2$/s\\
D$_T$ (drift field 1--10 kV/cm)~\cite{PhysRevA.20.2547,Derenzo:1974ji} &  $\sim 13-18$ cm$^2$/s\\
\hline
\end{tabular}
\end{minipage}
\end{table}

\subsection{Scintillation}
\label{sec:scint}
A charged particle crossing liquid argon causes ionization and excitation of the argon atoms, yielding free electrons and the emission of scintillation light. Scintillation light is observed~\cite{jortner:4250,Heindl:2010zz} in the VUV range in a band centered at 127 nm with a FWHM of about 10 nm, and is attributed to the emission of excited diatomic molecules (excimers). Transitions from the low excited singlet $^1\Sigma^+_u$ and triplet $^3\Sigma^+_u$ molecular states  to the ground state have been measured~\cite{Hitachi:1983zz} to have decay times of 7 ns and 1.6 $\mu$s, respectively; these decay times are independent of the density of ionization. Instead the intensity ratio of singlet to triplet molecular states depends on the ionization density~\cite{Hitachi:1983zz}: 0.3 for 1 MeV electrons and 3 for fission fragments.
LAr scintillation has also been observed in the near infrared~\cite{Bressi2000254,Buzulutskov2011}, due to transitions among the first excited atomic states. 

Ionization recombination takes place soon after the initial ionization charge around the trajectory of the ionizing particle is produced, depending on the ionization density and the value of the electric field in LAr. It effectively transforms ionization into scintillation. Excimers form~\cite{Doke1990617} through collisions between either excited atomic states and other atoms in the ground state or ionized atoms and ground state atoms in a sequence of steps involving recombination with electrons. About 70\% of the scintillation induced by 1 MeV electrons without any applied electric field comes from the recombination process, as measured in Reference~\cite{PhysRevB.17.2762} by simultaneously detecting a decrease in scintillation intensity and an increase in collected electrons while raising the intensity of the electric field.

The average energy required for the production of one photon in LAr has been estimated~\cite{Doke1990617,Doke1988291}, for 1 MeV electrons with no electric field, 
to be W$_{ph}$=25 eV, which is comparable to that for a NaI(Tl) crystal. These properties make LAr an excellent scintillator. For an electric field of 300 V/cm, the reduction in scintillation is about 35\%~\cite{Cennini:1999ih}.

The first measurements~\cite{Ishida1997380} of the attenuation length of LAr scintillation light in LAr itself yielded $\lambda_{att} = 66 \pm 3$ cm. Researchers suggested~\cite{Ishida1997380} that this attenuation is due to Rayleigh scattering in LAr, which was calculated~\cite{Seidel:2001vf} to be $90 \pm 32$ cm 
at the peak scintillation wavelength of LAr. Density fluctuations of LAr would greatly enhance Rayleigh scattering. Recently the attenuation of UV light in LAr, 
wavelength resolved in an interval from 118 to 250 nm, was measured~\cite{Neumeier:2012cz} to be $\lambda_{att}$(127 nm)=140 cm, with an error of $\sim 30\%$.
This value is marginally compatible with the previous measurement~\cite{Ishida1997380}. For large scale detectors employing LAr scintillation light, this is a critical quantity that deserves further investigation.

\subsection{Ionization}
\label{sec:ioniz}
The average energy required to produce an electron-ion pair in LAr is W$_i=23.6\pm 0.3$ eV~\cite{Miyajima:1974zz,Miyajima:1974zy}, a value that comes from the saturated collected charge from $^{207}$Bi conversion electrons at high electric fields. This value is smaller than that obtained in gaseous argon (W$_g$=26.4 eV~\cite{PhysRev.90.1120}), indicating the presence of a conduction band in the condensed state~\cite{Miyajima:1974zz}. 

Charge recombination measurements in LAr on low energy ($\lesssim 1$ MeV) electrons, as a function of the electric field, have been reported~\cite{Shibamura1975249,Scalettar1982,Thomas1987,Aprile1987519}.
Recombination effects for muons and protons for low drift electric fields (0.2--0 .5 kV/cm) have been measured  as a function of the stopping power $\langle dE/dx \rangle$ by the ICARUS 3 t and T600 detectors~\cite{Cennini:1994ha,Amoruso:2004dy}.

Different recombination models have been proposed~\cite{Jaffe1913,Onsager1938,Thomas1987}. 
Following Reference~\cite{marshall:232}, the Jaffe~\cite{Jaffe1913} ``columnar'' model
is approximated by the same expression as the so-called Birk's law describing the quenching effects in scintillators, Q/Q$_0$=(1+k$_E$/$\mathcal{E}$)$^{-1}$, where Q is the collected charge, Q$_0$ the initial ionization charge, $\mathcal{E}$ is the electric field and k$_E$ is a constant to be obtained from a fit to data. This expression does not seem to adequately fit the data of Ref.~\cite{Scalettar1982} over the full range of drift fields,
but it has been successfully used for high drift fields~\cite{Shibamura1975249,Aprile1987519,Scalettar1982}.

An explicit dependence on the stopping power $\langle dE/dx \rangle$  and a normalization constant have been introduced by Reference~\cite{Amoruso:2004dy}
\[
        Q=A \frac {Q_0} {1+k/\mathcal{E}\langle dE/dx \rangle}
\]
where A and k are parameters to be fit. This phenomenological expression, when fit to the ICARUS 3 t data~\cite{Cennini:1994ha} for drift fiels of 0.2--0.5 kV/cm,  yields~\cite{Amoruso:2004dy} A$= 0.800\pm 0.003$, and k$=0.0486\pm 0.0006$  kV/cm g/(cm$^2$ MeV). For minimum ionizing particles, it predicts a remaining charge after recombination equal to 70\% and 75\% of the initial ionization charge for electric fields of 0.5 kV/cm and 1 kV/cm, respectively. In a field of 0.5 kV/cm, an ionization charge of $\sim10$ fC/cm of track length is expected for minimum ionizing particles (Table~\ref{TabLAr_ph}).

The so-called box model~\cite{Thomas1987} predicts
\[
        Q=Q_0 \frac {1} {\xi} ln(1+\xi) 
\]
where $\xi$ is a parameter that is proportional to the ionization charge density and inversely proportional to the electric field. This expression has been successfully used~\cite{Thomas1987,Thomas1988} to fit the dependence of charge recombination for low energy electrons as a function of an electric field up to 10 kV/cm.

Given the importance of the recombination phenomenon for liquid argon, which directly affects the discrimination of particle types on the basis of their stopping power and the achievable energy resolution, additional measurements of this effect would be welcome.

The intrinsic energy resolution of 1 MeV electrons in LAr has been measured~\cite{Aprile1987519}, through the detection of the ionization charge, to be 2.7\% (FWHM). This value is still a factor 2 poorer than the the Poisson statistical limit and a factor 7 poorer than the Fano limit~\cite{Doke:1976zz}. The loss of resolution has been explained~\cite{Thomas1988} by the statistical fluctuations in the number of soft $\delta$ rays along the electron track; these $\delta$ rays suffer from larger recombination than do higher energy electrons.

\subsection{Drift and Diffusion}
\label{sec:drift}
The drift velocity of electrons in LAr depends primarily on the electric field strength~\cite{Shibamura1975249}. A weaker dependence on the temperature has been measured~\cite{Walkowiak:2000wf}. The addition of small amounts of impurities to LAr causes the drift velocity to increase~\cite{Shibamura1975249}. Precise measurements of the drift velocity as a function of the drift electric field and the temperature of the liquid have been performed~\cite{Walkowiak:2000wf} for electric fields $\mathcal{E}$ and temperatures T in the ranges between 0.5 kV/cm $\le \mathcal{E} \le$ 12.6 kV/cm and 87 K $\le T \le$ 94 K, respectively. A relative accuracy of $\lesssim 2\%$ was obtained for fields $< 4$ kV/cm. The mean value of the temperature dependence of the drift velocity $v_d$ has been measured to be $\Delta v_d/\Delta T v_d = (-1.72 \pm 0.08)\%/K$. The ICARUS collaboration~\cite{Amoruso:2004ti} has performed measurements of the drift velocity for electric fields as low as 60 V/cm and up to 1 kV/cm at T=89 K. These values agree well with those from Reference~\cite{Walkowiak:2000wf} in the overlap region.
 
The drift velocity of electrons increases less than linearly for electric fields above 100 V/cm, still there is considerable gain at higher fields. It reaches 1.55 mm/$\mu$s at 0.5 kV/cm for a temperature T=89 K, and it increases by 30\% by doubling the electric field intensity from 0.5 to 1 kV/cm, and again by 30\% from 1 to 2 kV/cm.

Knowledge of the diffusion coefficient D is especially important for long drift times because it directly affects the accuracy of the drift time measurement and the transverse smearing of the reconstructed tracks. The spread of an isolated group of electrons increases with drift time t as $\sigma = \sqrt{2 D t}$.
For gases, the diffusion coefficient in a direction parallel to an electric field (D$_L$) differs from that in the transverse direction (D$_T$)~\cite{Lowke1969zz}. For argon gas D$_L$ is substantially lower than D$_T$.

Diffusion measurements of electrons in LAr are scarce. An initial measurement~\cite{Derenzo:1974ji} yielded D$_T \simeq 13$ cm$^2$/s for a drift field of 2.7 kV/cm.
The mean agitation energy $\langle \epsilon \rangle$ of electrons in LAr was estimated~\cite{PhysRevA.20.2547} through the ratio $\langle \epsilon \rangle = eD/\mu$ of the diffusion coefficient to mobility $\mu$ by measuring the lateral spread of a group of electrons in a parallel plate ionization chamber operated with electric fields between 2 and 10 kV/cm. The derived values for D$_T$ lie in a range between $\sim 15 - 18$ cm$^2$/s. The value of D$_L$ for low drift fields, $\sim 100 - 350$ V/cm, was determined~\cite{Cennini:1994ha} by an analysis of the rise time of the electron collection signal; the result, D$_L =4.8 \pm 0.2$ cm$^2$/s, agrees reasonably well with the Einstein relation eD/$\mu =$kT for thermal electrons.

\subsection{Electron Attachment to Impurities}
\label{sec:purity}
Following Reference~\cite{Bakale1976}, in presence of electronegative impurities, the concentration of free electrons [e] in LAr decreases exponentially with time according to
\[
        \frac {d[e]} {dt} = - k_s [S] [e] \quad \Rightarrow \quad [e]=[e_0] e^{-k_s [S] t}=[e_0] e^{-t/\tau_e}
\]
where [S] is the concentration of electronegative impurities, k$_s$ is an electron attachment rate constant and $\tau_e = 1/(k_s [S])$ is the electron lifetime. Knowledge of the effect of oxygen is particularly important because it is one of the main contaminants in commercial LAr. The attachment rate constants to oxygen have been measured~\cite{Bakale1976} as a function of the electric field strength. For electric fields of less than a few 100 V/cm, the results is k$_{O_2} = 9 \times 10^{10}$ L mole$^{-1}$ s$^{-1}$ = 3 ppm$^{-1}$ $\mu$s$^{-1}$, which corresponds to $\tau_e (\mu s) \simeq 300/$[O$_2$] (ppb). For increasing electric fields, the attachment rate constant decreases, as already observed in the early Reference~\cite{Swan196374}. At an electric field of 1 kV/cm, $\tau_e (\mu s) \simeq 500/$[O$_2$] (ppb). 

The mean free path length $\lambda$ of electrons in LAr was measured~\cite{Hofmann:1976de} to be inversely proportional to the O$_2$ concentration and linearly proportional to the drift electric field $\mathcal{E}$ for small values of the electric field: $\lambda = \alpha \mathcal{E}$/[O$_2]$, where $\alpha = (0.15 \pm 0.03)$ ppm cm/(kV/cm).  For low electric field electron mobilities, namely $\sim 500$ cm$^2$ V$^{-1}$ s$^{-1}$~\cite{Buckley:1988qx}, this value is equivalent to the above expression for the electron lifetime. 

\subsection{LAr-Vapor Interface}
\label{sec:doublephase}
The characteristics of electron emission from the liquid~\cite{Dolgoshein:1970,Dolgoshein:1973} have been experimentally studied~\cite{Gushchin:1982,Borghesani1990481}. These investigtions confirmed the existence of a negative energy level of conduction electrons in LAr, 
of about $0.1-0.2$ eV~\cite{Gushchin:1982,Borghesani1991615,Schmidt1984xx}. The emission of the electrons can be described by the Schottky model of electric field enhanced thermoionic emission.

\section{CHALLENGES FOR LARGE LIQUID ARGON DETECTORS}
\label{sec:RandD}
\begin{figure}
\includegraphics[width=0.99\linewidth]{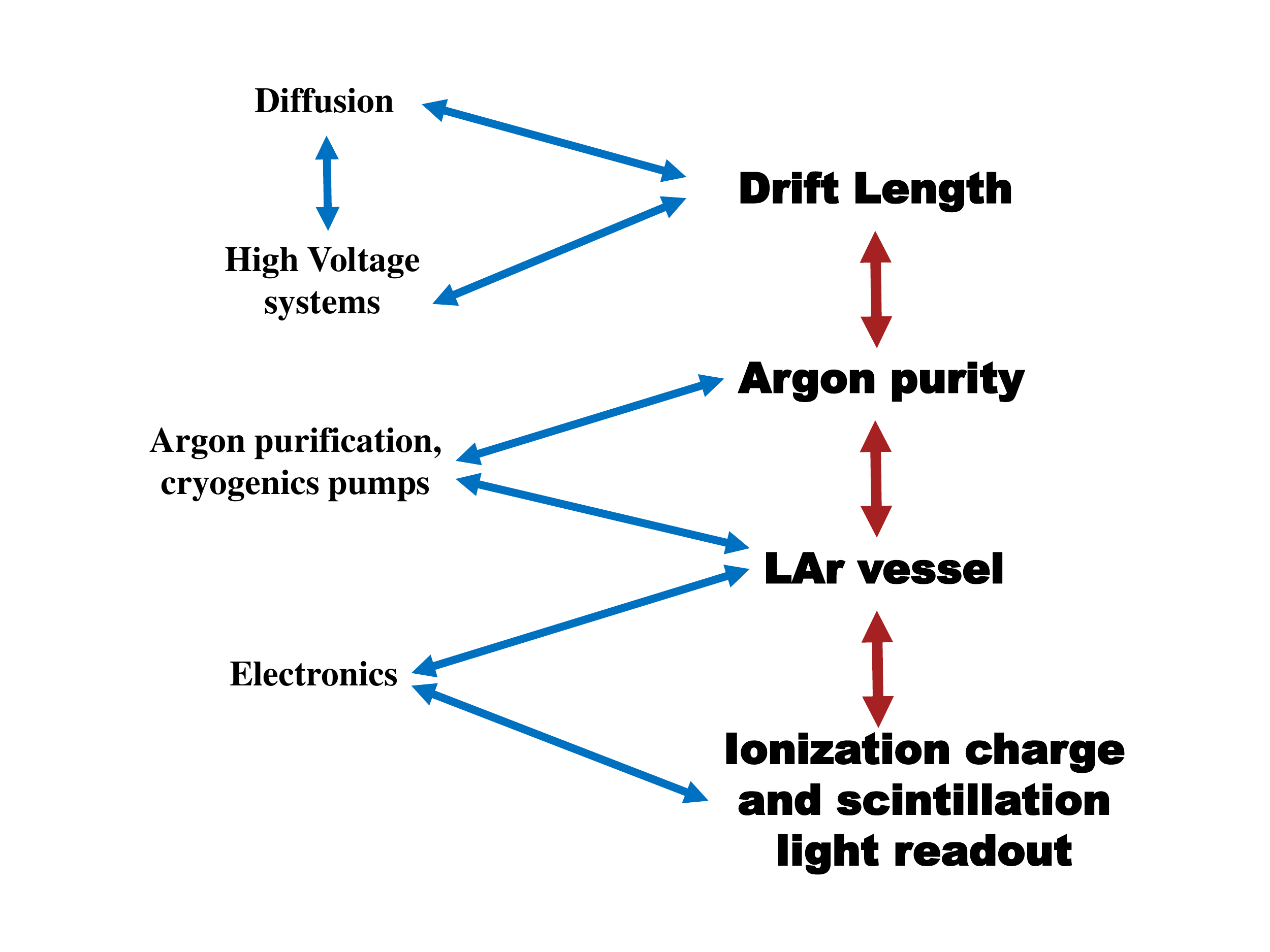}
\caption{Main design topics for LAr TPCs.}\label{fig:RandD}
\end{figure}

Operation of a tens of kilotonnes LAr TPC detector could answer many fundamental physics
questions. Doing so would require more than a factor of 10 increase in size from the present
ICARUS T600 detector. The main factors that have been considered in extrapolations to larger
detectors include larger cryogenic vessels, longer drift paths to minimize the number of electronic
channels per unit volume, lower cost electronics (possibly located next to the readout elements and
immersed in LAr), novel readout methods for ionization charge and scintillation light, and new
schemes for the generation of high voltages in LAr. The achievable purity of LAr remains a main
design parameter for a LAr TPC. Figure~\ref{fig:RandD} summarizes these main design considerations and the
relationships among them. Monitoring and calibration of the detector performance parameters,
such as the purity of the argon and the uniformity of the drift region, are also of primary importance.

\subsection{Cryogenic Vessels}
 The envisioned future detectors require large cryogenic vessels ($\gtrsim 10,000$ m$^{3}$), suitable for underground construction and operation, that are sufficiently tight to keep electronegative impurities within a few tens of ppt and with low enough thermal losses to afford the costs of long term cryogenic operation.
Traditionally high purity LAr vessels were built as stainless-steel vacuum insulated dewars, in which the inner vessel is also evacuable. Such devices, if built with standard techniques, are not scalable to the required dimensions of the envisioned LAr detectors. In Section~\ref{sec:designs_evac}, we describe a design that aims to maintain the features of vacuum insulation and evacuable vessels for very large containers.

The ICARUS T600 vessel utilizes a cryostat made of two identical, adjacent, evacuable aluminum vessels, each of which contains 300 t of LAr that are thermally insulated by evacuated honeycomb panels~\cite{Amerio:2004ze,Rubbia:2011ft}. The detector is cooled by pressurized liquid nitrogen, which circulates in a thermal shield placed between the insulation panels and the aluminum vessels. 

Very large cryogenic vessels would hardly be evacuable. In Section~\ref{sec:nonevacuable}, we show that this limitation does not necessarily preclude the achievement of the necessary LAr purity. Moreover, a high volume to surface ratio is advantageous for keeping the detector in cryogenic conditions, for allowing passive insulation, and for maintaining the purity of the argon. Industrial tanks used to contain liquefied natural gas (LNG, which comprises more than 90\% CH$_{4}$ )~\cite{Fulford1988810}, up to volumes of 200,000 m$^{3}$, are built as nonevacuable nickel steel tanks; perlite is used as thermal insulation. 
Several groups~\cite{Mulholland:2002,Rubbia:2004tz,Bartoszek:2004si} have proposed the use of such vessels for the containment of LAr, given that the boiling points of LAr and CH$_{4}$ are similar (87.3 and 111.6 K, respectively) and the latent heat of vaporization per unit volume 
is the same for both liquids within 5\%.
Thermal losses of $\sim$ 5 W/m$^{2}$ are attainable with passive insulation methods (i.e. with $\sim 1.5$ m thick perlite insulation). The resulting boil-off of LAr, of only $\sim 0.05\%$/day for a 100 kt LAr vessel, would then be recondensed and returned to the cryostat.

A different solution, again borrowed from the field of LNG technology, has been adopted for the LAr detector of the LBNE project~\cite{LBNE_LAr:2012zz} at Fermi National Accelerator Laboratory (FNAL). This system employs a commercial membrane cryostat technology, used in LNG tanker ships~\cite{FS1993772}, that involves a  $\sim 1$ mm thick corrugated stainless-steel liner to contain the cryogenic liquid and relies  on the surrounding rock to provide mechanical support.  
A secondary thin aluminum membrane, serving as secondary containment, is embedded in a 0.8 m thick passive insulation of solid, reinforced polyurethane foam.
The heat input from the surrounding rock is expected to be $\sim 7$ W/m$^2$, which causes  continuous evaporation of the argon because the cryostat is not directly cooled in these designs. Fluid dynamics simulations show the onset of convective currents throughout the entire LAr bath, with speeds of a few cm/s, that help limit the temperature gradient across the whole volume to much less than 0.1 K, so the variations of the drift velocity is less than 0.1\% (see Section~\ref{sec:drift}). The evaporated argon is recovered, recondensed by LN$_2$ refrigerators, purified and returned to the cryostat.

\subsection{HV Generation and Feedthroughs}
The development of HV systems is crucial for LAr TPCs, which need a stable and uniform drift field over extended LAr volumes. 
Typically, a drift field of 500 V/cm is utilized. There are clear advantages in operating a LAr TPC at a higher drift electric field, 
including shorter charge collection times, which reduce the attenuation of ionization electrons due to electronegative impurities; 
lower recombination of the initial ionization charge, particularly for heavily ionizing particles; 
and lower attachment cross sections of electrons to O$_2$ impurities, as mentioned Section~\ref{sec:purity}.
However,  a high drift field may increase the diffusion process and lead to very high voltages on the cathode. 
It may also require unacceptably high voltages on the charge readout structure to ensure transparency for the collection of the ionization charge.

The HV configuration chosen for ICARUS T600, described in Section~\ref{sec:TPCdetector}, allows access to the HV power supply for possible repairs, as well as continuous monitoring of the voltage from the current flowing in the resistive voltage divider. The design of the HV feedthrough, operated at a 
maximum voltage of 150 kV, is in principle scalable to higher voltages, as long as suitable HV power supplies and cables are available.

An alternative solution~\cite{Horikawa:2010bv,Ereditato:2013xaa,Badertscher:2012dq} is to build a Cockcroft-Walton voltage multiplier immersed in LAr;
the different stages of the multiplier would be directly connected to the field shaper electrodes and to the cathode, 
thereby avoiding the need for a resistive voltage divider. 
Because in this case a negligible current is drawn, the voltage multiplier can be operated at a relatively low frequency charging voltage, 
well outside the relevant bandwidth of the readout electronics.
In principle, this concept allows one to reach HV values above the range of commercially available power supplies.

 A proof of principle is provided in Reference~\cite{Badertscher:2012dq}, which describes, in detail, the design, characteristics and operation of a voltage multiplier with 30 stages that is driven by a 50 Hz AC charging voltage supply up to a maximum driving voltage of V$_{pp}=2.0$ kV peak-to-peak. This device was successfully operated up to a voltage of $\approx 35$ kV. 

Recently, a voltage multiplier~\cite{Ereditato:2013xaa} with 125 stages, driven by a maximum AC voltage of V$_{pp}=4.0$ kV, peak to peak, 
was stably operated at a voltage of up to $\sim120$ kV. Discharges encountered at higher voltages are presently under investigation.

Voltage multipliers with a large number of stages are affected by a so-called ``shunt'' capacitance becuse of the capacitance of the diodes and the stray capacitance determined by the geometry of the circuit. This factor causes the output voltage to increase less than linearly with the stage number, even with no current load~\cite{Horikawa:2010bv,Badertscher:2012dq}. Designs for up to megavolt systems have been discussed in Reference~\cite{Horikawa:2010bv}.

\subsection{LAr Purity and Purification Methods}
\label{sec:purify}
The purification of LAr, down to electronegative impurity concentrations of a few tens of ppt, and the maintenance of such conditions over many years are key 
to the successful operation of large LAr TPCs. Continuous recirculation and purification of LAr are necessary in such detectors, as demonstrated by ICARUS.
Argon purity remains an important R\&D subject, focusing primarily on achieving purity in nonevacuable vessels, on argon purification techniques, on the qualification of materials used in detector construction, and on monitors of argon purity.

\subsubsection{Nonevacuable vessels}
\label{sec:nonevacuable}
In many designs of large LAr TPCs, the cryogenic vessel is nonevacuable and requires the removal of air before it can be filled with LAr. 
Preliminary tests have been conducted on small tanks~\cite{Jaskierny:2006sr,Curioni:2010gd} of a few m$^3$ in the gas phase; argon gas
was injected at the bottom of the tank to smoothly displace the air exiting at the top.

Large-scale tests at FNAL with a 30 t LAr tank (Liquid Argon Purity Demonstrator project~\cite{Rebel:2011zzb}) are under way. 
The purification of the tank, starting from air, proceeds through three main steps: 
an initial purge with argon gas, gas recirculation while the tank is heated to $50^\circ$ C to remove water, and LAr filling and recirculation.
Preliminary results~\cite{Tope:2011} are very encouraging; they show electron lifetimes of $\sim3$ ms.

\subsubsection{Purification methods}
The ICARUS collaboration~\cite{Amerio:2004ze} has used commercial Oxisorb filters in combination with molecular sieves for argon purification. 
Regenerable custom-made purification cartridges that employ reduced copper as the purification agent were first used~\cite{Doe1987170} for gas phase argon 
and were later successfully operated~\cite{Curioni:2009rt,Badertscher:2010zg} directly in LAr. In particular activated-copper-coated alumina granules have been employed at FNAL~\cite{Curioni:2009rt}. Such cartridges are easily regenerable at about $250 ^{\circ}$C in a 95:5 mixture of Ar:H$_{2}$ gas. Combined with an upstream molecular sieve filter, of type 4A, electron lifetimes of $\sim$10 ms, at the upper limit of the purity monitor range, have routinely been obtained~\cite{Curioni:2009rt}.

\subsubsection{Qualification of materials}
A facility has been created at FNAL~\cite{Andrews:2009zza} to test a number of materials commonly used in detector construction. No effect on the electron lifetime was measured when these materials were immersed in LAr, but when they were positioned in the warmer region of the vapor phase (at $\sim200$ K), the electron lifetime strongly decreased; this reduction in electron lifetime correlated with an increase of water concentration in the vapor phase. 
Water may be responsible for the decrease of the lifetime, and water concentrations in liquid phase at the 10 ppt level may affect the electron lifetime~\cite{Andrews:2009zza}. 

\subsubsection{Monitors of argon purity}
An effective way to monitor the attenuation of the ionization charge, which occurs because of electronegative impurities, is to make a three-dimensional reconstruction 
of cosmic muons in a TPC. From this reconstruction, one can measure the collected charge per unit length versus drift distance. 
However, it is important to have devices that can perform such measurements both in real time and without relying on the full functionality of the TPC. 
The purity monitor described in Reference~\cite{Carugno:1990kd} is suitable for electron lifetimes from a few microseconds to a few milliseconds. 
It was further optimized for use in ICARUS T600~\cite{Amerio:2004ze}.

A novel monitoring and calibration system that exploits UV laser ionization in LAr has been devised~\cite{Rossi:2009im}. 
Laser generated tracks can be used to measure the drift velocity, the uniformity of the drift field and to determine the electron lifetime.

UV laser ionization of LAr for TPC calibration was first studied~\cite{Sun:1996tn} by use of a pulsed UV
Nd:YAG laser, which operated on the fourth harmonic with a wavelength of 266 nm, corresponding
to 4.66 eV photons. LAr ionization occurs as a multiphoton process wherein a two-photon, near-resonance excitation of an argon atom 
is followed by absorption of another photon, causing ionization. 
This mechanism was subsequently studied in Reference~\cite{Badhrees:2010zz}, which identifies an important disagreement with the 
previous measurement~\cite{Sun:1996tn} of the two-photon excitation cross section; this discrepancy is not yet understood.

The decay time of the slow component (t$_{slow}$) of LAr scintillation light is very sensitive to the presence of impurities in LAr; 
its dependence on the concentration of impurities has been studied in References~\cite{Acciarri2010P06003} and~\cite{Acciarri:2008kx} 
for contamination by N$_2$ and O$_2$, respectively. Measurements of t$_{slow}$ have been used as a simple monitor of impurities, 
with a sensitivity down to 0.1 ppm~\cite{Amsler:2010xx}.

\subsection{Detection of Ionization Charge and Scintillation Light}
\label{sec:readout}
The choice of the readout method for the ionization charge released in LAr and the approach used for the detection of scintillation light play essential roles in the design of a large LAr TPC. 

\subsubsection{Charge readout in liquid phase with wire planes}
\label{sec:wireread}
The wire readout technique~\cite{Gatti:1979} offers the possibility to acquire three independent views of events in the very stable liquid phase conditions. The LAr density
at the normal boiling point changes by -0.44\%/K; temperatures are typically stable to much lower than 1 K. 

Direct scaling from the ICARUS T600 design to much larger dimensions is not feasible. The mechanical and electrical properties of the wires, namely tension, sagging, the possibility that vibrations will induce microphonic noise, electrical resistance, and capacitance, must be checked and optimized. The total input capacitance to the amplifiers scales with the wire and signal cable lengths and may become a limiting factor in the design due to the worsening of the signal-to-noise ratio S/N~\cite{Radeka:2011zz}. Typical capacitances are $\sim 20$ pF/m and $\sim 50-100$ pF/m for wires and signal cables, respectively. Input capacitances exceeding 400 pF are expected for 10 m long wires, when preamplifiers are located outside a large cryostat.

For large capacitance detectors, the dominant noise is the series noise (voltage noise) of the first transistor of the amplifier; it  increases linearly with the input capacitance.
In the case of warm electronics, placed outside the cryostat, JFETs are usually employed as the first stage of the amplifiers because of their sufficiently high transconductance and their extremely small parallel noise (current noise)~\cite{Radeka1988217,Angeli:2009zza}. 
The equivalent noise charge (ENC) of the JFET can be expressed as~\cite{Radeka:2011zz,Radeka1988217}
\[
        ENC^2=e_{sn}^2C_{in}^2/t_p
\]
where $e_{sn}^2=4kT(2/3)/g_m$ is the series noise, $C_{in}$ is total input capacitance, equal to the sum of the wire, signal cable and input transistor capacitances, and $t_p$ is the shaping time.
The basic design of preamplifiers with multiple JFETs at the input stage, connected in parallel to increase the transconductance $g_m$, was first employed by the ICARUS collaboration~\cite{Amerio:2004ze} and successfully adopted by other groups~\cite{Badertscher:2008rf,Badertscher:2013wm}. The ICARUS T600 amplifier has an ENC of $\sim1250$ electrons at  $C_{in} = 450$ pF~\cite{Angeli:2009zza}, providing a S/N $\sim 10$ in the case of minimum ionizing particles for 3 mm wire spacing.

In the case of long wires, it is important to use a low-resistivity wire material to limit the input thermal noise to the amplifier~\cite{Rescia20091932}. 
For stainless-steel wires, the noise contribution due to the resistivity of the wire becomes important for wire lengths $\gtrsim 5$ m.

In a novel concept adopted for the LBNE project~\cite{LBNE_LAr:2012zz,Radeka:2011zz}, the LAr volume is filled with a 3D array of TPC cells, with a maximum wire length of $\sim 10$ m, and specifically designed front-end electronics operate in LAr next the wire ends, eliminating the need for long readout cables. 
Application-specific integrated circuits (ASICs) based on CMOS technology for front-end applications for LAr TPCs, operating in LAr inside the cryostat, have recently attracted increased attention~\cite{Autiero:2009xy,DeGeronimo:2011zz}. Standard test methods are being applied the verify the lifetime of these components~\cite{Radeka:2011zz,LBNE_LAr:2012zz}, which will be required to operate for more than 20 years.

The MicroBooNE experiment~\cite{MicroBooNETDR} has developed a 180 nm CMOS ASIC~\cite{Chen2012C12004} to be operated in LAr. This ASIC has a power consumption of only $\sim 6$ mW per channel. Test results show an ENC that decreases from $\sim 1200$ electrons at 293 K to fewer than 600 electrons at 77 K with 150 pF input capacitance. These results demonstrate the possibility of achieving an ENC smaller than 1,000 electrons for a 10 m long wire with front-end electronics 
located in LAr in close proximity to the wire end.

The integration of an analog front-end amplifier with a digitizer in the same ASIC, operated at LAr temperature, 
is presently being developed~\cite{DeGeronimo:2011zz}; the use of both electrical and optical cryogenic digital data links, 
for the transmission of the signals to the outside of the cryostat, is being investigated~\cite{Liu:2012zzc}. 
Such a solution would substantially reduce the number and complexity of cryogenic feedthroughs.

\subsubsection{Scintillation light detection}
The ICARUS T600 detector~\cite{Amerio:2004ze} uses tetraphenyl-butadiene (TPB)~\cite{Burton:73} coated PMTs to shift the 127 nm LAr scintillation light to the visible region. Recent measurements of the visible reemission spectrum and fluorescence efficiency of TPB, using a TPB film produced by vacuum evaporation, have been
reported~\cite{Gehman:2011xm}. 
Under the assumption of a Lambertian angular distribution of the reemitted light from both sides of the TPB film, $\sim1.2$ photons are emitted, in a band centered at $\sim 420$ nm, for every 127 nm photon absorbed.

The MicroBooNE experiment~\cite{MicroBooNETDR} uses, as wavelength shifter, acrylic plates coated with a solution of TPB and polystyrene; the plates are placed in front of the PMTS. A strong decrease of the fluorescence efficiency has been observed~\cite{Chiu:2012ju} for plates exposed to common flurescent lights and sunlight, because of their UV component. 
Large area light collection systems are needed for large LAr TPCs. One solution~\cite{Bugel:2011xg} uses acrylic light guides coated with a solution of polystyrene and TPB. Multiple bars are bent to adiabatically guide the light to a single PMT, thereby efficiently using the active area of the PMT. 

\subsubsection{Charge collection and amplification in double phase}
\label{sec:doublephaseread}
A charge imaging LAr TPC operated in double phase, denominated Large Electron Multiplier Time Projection Chamber (LEM TPC), was first suggested~\cite{Rubbia:2004tz} to improve imaging capabilities and somewhat compensate for charge attenuation in very long drift paths. Ionization electrons are drifted towards the liquid surface at the top of a vertical cryostat, are then extracted in pure argon gas, and are amplified by thick macroscopic GEMs (Large Electron Multipliers, also known as THGEM). 

A very active R\&D program is being conducted on THGEMs~\cite{Breskin:2008cb}, which are macroscopic hole charge multipliers manufactured with standard PCB techniques. They have been successfully operated in noble gases, without any added quenching gas, and in double-phase argon detectors~\cite{Buzulutskov:2011de}.

The first LEM TPC prototype was successfully operated~\cite{Badertscher:2008rf} with a double stage THGEM of $10\times 10$ cm$^{2}$. The amplified charge was readout via two orthogonal coordinates by use of the induced signal on the upper segmented plane of the top THGEM and the collection signal on a segmented anode.
 
In an improved design~\cite{Badertscher:2010zg}, to decouple the mechanism of charge amplification in the THGEM from the charge readout, the amplified charge is drifted to a two-dimensional projective readout anode, which provides two independent views with 3 mm spatial resolution. 
An effective $\sim30$-fold amplification of the released ionization charge has been achieved~\cite{Badertscher:2010zg} with a single $10\times 10$ cm$^{2}$
THGEM configuration, operated with an electric field of $\sim35$ kV/cm, without any degradation in resolution; minimum ioninzing particles have been detected 
with an S/N larger than 200.

The readout of a large area detector would be realized by an assembly of many independent readout units, each still of a relatively large area of the order of 1 m$^2$. Recently, a LEM TPC with a readout unit of $76\times40$ cm$^2$ was successfully tested~\cite{Badertscher:2013wm} with an effective $\sim14$-fold charge
amplification.

\subsubsection{Charge readout through light}
Signals from an $^{55}$Fe source were observed~\cite{Lightfoot:2008ig} with good energy resolution in a double phase argon setup through detection of the secondary scintillation light produced by the ionization electrons extracted from the liquid and drifted into the holes of a THGEM. This device operated with electric fields $\sim 20$ kV/cm in pure argon gas. The scintillation light was detected by a single 1 mm$^{2}$ silicon photomultiplier coated with TPB. The application of this technique to large LAr TPCs could be attractive, but it would require substantial R\&D on low cost photosensors.
The claim~\cite{Lightfoot:2008ig} that secondary scintillation light is also observed when a THGEM is operated while immersed in LAr, with fields $\sim 60$ kV/cm, 
has yet to be confirmed.

\subsection{Long Electron Drifts}
\label{sec:drifts}
Electron drifts of at least a couple of meters are necessary for the realization of a realistic 10 kt scale LAr detector, even if it is built in a modular way. 
A 5 m drift device~\cite{Ereditato:2013xaa} has been constructed for a full scale demonstration of very long drifts, directly addressing the issues of the generation 
of very high voltages (up to 500 kV) and the achievable S/N. In a first test~\cite{Ereditato:2013xaa}, performed in 2012, cosmic muon tracks were observed for drift distances up to 5 m.


\section{TOWARDS LARGE LAr DETECTORS}
\label{sec:designs}
\begin{figure}
\includegraphics[width=0.98\linewidth]{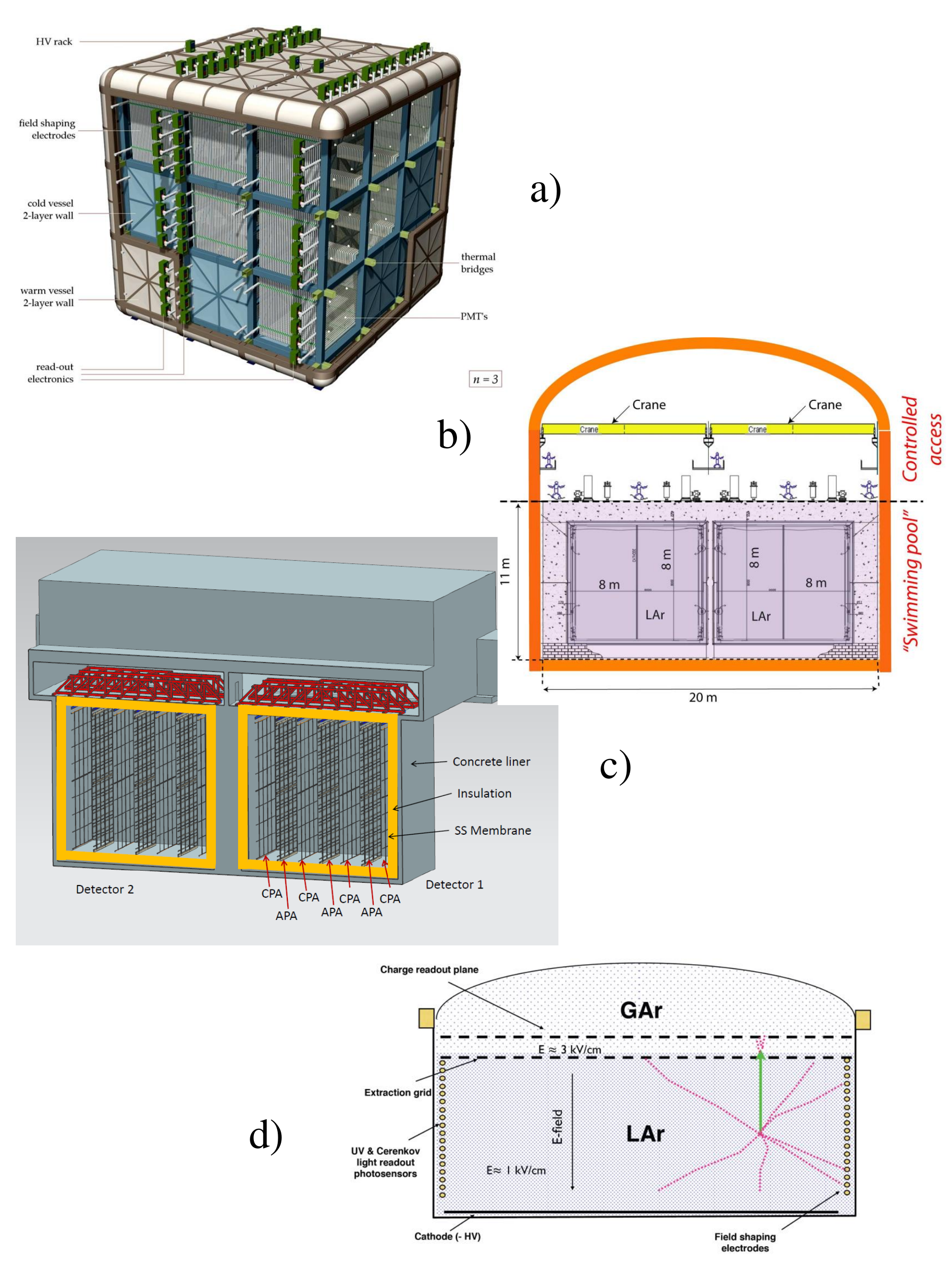}
\caption{
Main design concepts for large LAr TPCs. 
a) An evacuable cryostat with vacuum insulation~\cite{Cline:2006st}. 
b) A nonevacuable cryostat with passive insulation, built from 5 kt modules extrapolated from ICARUS~\cite{Angeli:2009zza}. 
c) A nonevacuable cryostat based on membrane cryostat technology, with immersed electronics distributed inside the
detector volume~\cite{LBNE_LAr:2012zz}.
d) A nonevacuable cryostat, based on the liquefied natural gas tank design, with vertical drift of the ionization
charge and subsequent amplification in pure argon gas at the top~\cite{Rubbia:2012jr}.
Abbreviations: APA, anode wire plane assembly; CPA, cathode plane assembly; HV, high voltage; SS: stainless-steel.
}\label{fig:bigLAr}
\end{figure}

Motivated by the necessity of large detectors for neutrino physics, proton decay and the observation of astrophysical neutrinos, several groups have recently proposed designs of large LAr detectors (up to $\sim$100 kt)~\cite{Cline:2001pt,Rubbia:2004tz,Bartoszek:2004si,Cline:2006st,Baibussinov:2007ea,Angeli:2009zza,Rubbia:2009md,LBNE_LAr:2012zz,Stahl2012xx}.
Figure~\ref{fig:bigLAr} summarizes the main concepts of these designs, which we discuss further in this section.

\subsection{Large Evacuable Vessels with Charge Readout in Liquid Phase}
\label{sec:designs_evac}
Reference~\cite{Cline:2006st} proposes scalable detectors, based on a three-dimensional cubic frame array immersed in a common LAr volume (see Figure~\ref{fig:bigLAr}a for a 3$^3$ configuration), that have a basic cubic cell of $\sim 200$ m$^3$ and a drift length of $\sim 5 $ m. Such a mechanical structure allows for the evacuation of the inner vessel, so as to preventively check its tightness, but at the expense of complicating the construction and assembly of the readout devices. The design assumes a double wall cryostat, with vacuum insulation around the cold vessel to optimize thermal insulation. 

\subsection{Large Nonevacuable Vessels with Charge Readout in Liquid Phase}
The MODULAr proposal~\cite{Baibussinov:2007ea,Angeli:2009zza} argues that a tens of kilotonnes LAr TPC would best realized with a modular set of $\sim 5$ kt independent units. The geometry is directly scaled from the ICARUS T600 detector, by about a factor 3 in each dimension, corresponding to a 4 m drift length  (see Figure~\ref{fig:bigLAr}b). The wire pitch is 6 mm to compensate for the higher input capacitance due to the longer wires. The vessel is nonevacuable, with 1.5 m thick perlite walls as passive heat insulation.

The LBNE LAr TPC~\cite{LBNE_LAr:2012zz} will consist, in its initial phase, of two stainless-steel membrane cryostats, each of which has a total mass of 9.4 kt of LAr 
and a fiducial mass of 5 kt. The concept is based on 2.3 m drift double TPC cells, with one double sided anode wire plane assembly (APA) in the middle and a cathode plane assembly (CPA) on each side (see see Figure~\ref{fig:bigLAr}c).  The TPC cells, 7 m high and 2.5 m wide in the beam direction, are repeated in the three dimensions within the cryostat; there are no constraints on the length of the signal cables thanks to the use of cryogenic electronics. The wire planes provide 3 independent views with a 4.5 mm wire pitch.

\subsection{Large Nonevacuable Vessels with Charge Readout in Double Phase}
A single 100 kt LAr module, denominated GLACIER, has been proposed (Figure~\ref{fig:bigLAr}d)~\cite{Rubbia:2004tz, Rubbia:2009md,Rubbia:2012jr}.
GLACIER has an almost totally active LAr mass. The vessel, in its original design, is a cylindrical stainless-steel 9\%-nickel tank based on industrial LNG technology;
its diameter is 70 m diameter and its height 20 m. The ionization charge is vertically drifted with an electric field of $\sim 1$ kV/cm, provided by an immersed Cockroft-Walton voltage multiplier, and collected and amplified in the argon vapor phase (see Section~\ref{sec:doublephaseread}). A 20 kton detector, with 20 m drift, 
was recently put forward as the initial phase of the LBNO proposal~\cite{Stahl2012xx} at CERN. Both a stainless-steel 9\%-nickel tank
and a membrane cryostat are presently being investigated as options for the LAr vessel.

\subsection{Test-Beam Setups and Intermediate Scale Detectors}
Test-beam setups and intermediate scale detectors are important because they integrate some
design aspects of large LAr detectors and complement the bench-type devices (Section~\ref{sec:RandD})
that address specific technological points. They enable physics measurements, the development of
simulation tools and reconstruction software, and assessments of the performance of LAr detectors
(i.e., electron/$\pi^0$ separation) in real working conditions. Here, we describe devices that have
recently been operated or are presently being built.

ArgoNeuT~\cite{Anderson:2012vc,Anderson:2011ce,Anderson:2012uk}, a 170 liter active volume detector, has been exposed to the NuMI
neutrino line at FNAL. It has collected neutrino interactions in the $0.5-10$ GeV range. In the
near future, the ArgoNeuT detector will be exposed to a low energy test beam~\cite{Adamson:2013/02/28tla} at FNAL to
study ionization charge recombination and particle identification capabilities.

The MicroBooNE detector~\cite{MicroBooNETDR}, with a $\sim60$ t fiducial LAr mass, is presently under construction
at FNAL and will be located on the Booster neutrino beam line. MicroBooNE will test some
of the important design aspects of large LAr TPCs, such as the use of cryogenic electronics in LAr,
the achievement of adequate argon purity in a large nonevacuable vessel, operation with a $\sim2.5$ m
drift length, and the use of foam as thermal insulation for the cryostat, directly measuring
the uniformity of temperature and drift velocity. This detector will collect more than 100,000
low energy neutrino interactions, will address cosmic ray backgrounds to proton decay searches,
and will be equipped with a trigger for supernova neutrino detection.

At the J-PARC facility, a 250-liter LAr TPC~\cite{Araoka:2011pw} has been exposed to a $500-800$ MeV hadron
beam, enriched in kaons. This test will provide particle identification capabilities in a LAr TPC
through the measurement of specific ionization versus residual range for pions, protons, and
kaons. It will also provide an estimate of efficiency and background for proton decay searches in
the charged kaon decay mode.

\section{CONCLUSIONS}
\label{sec:conclusions}
The use of LAr TPCs as detectors for neutrinos, both from accelerator beams and astrophysical
sources and for rare event searches, has been proposed since the 1970s. A vigorous R\&D effort
from the ICARUS Collaboration has culminated in the successful operation of the T600 detector
at LNGS, Italy.

Several large LAr TPC detectors, based on somewhat different approaches, have been proposed
during the past 10 years. Engineering studies have been performed for large cryogenic vessels;
they have mostly attempted to adapt industry standards from LNG technology. Experimental
tests on LAr purification have been performed, beginning from nonevacuable vessels. Substantial
developments in the realization of cryogenic electronics have given rise to designs of large vessels
with electronics distributed throughout the volume. A renewed interest in double phase operation
has prompted the development of novel detectors with amplification of the ionization charge in
the gas phase. More efficient and cost effective solutions have been proposed for the detection
of scintillation light. The wealth of new developments has allowed us to create comprehensive
designs and make reliable cost estimates. The exposure of small or intermediate scale LAr TPCs
to hadron and neutrino beams, in addition to their specific technical and physics goals, has helped
us develop new communities interested in LAr technology, with focused efforts in software and
event reconstruction.
\section*{Acknowledgments}
I gratefully acknowledge my many colleagues at ETH Zurich, FNAL, CERN, and KEK for
enlightening discussions on liquid argon technology. In particular, I thank Bruce Baller, Stephen
Pordes, Andr´e Rubbia, and Franco Sergiampietri. Work on this review was supported by the U.S.
Department of Energy under contract DE-AC02-07CH11359.
\bibliography{LArbib}
\bibliographystyle{arnuke_revised}
\end{document}